\documentclass[%
 reprint,
 showpacs,
 showkeys,
 preprintnumbers,
 amsmath,amssymb,
 aps,
 pra,
  longbibliography,
 ]{revtex4-1}

\usepackage[breaklinks=true,colorlinks=true,anchorcolor=blue,citecolor=blue,filecolor=blue,menucolor=blue,pagecolor=blue,urlcolor=blue,linkcolor=blue]{hyperref}
\usepackage{graphicx}
\usepackage{xcolor}

\begin{document}

\title{Unice cogito, ergo quantum sum\\ (I think uniquely, therefore I am quantum mechanical)}

\author{Karl Svozil}
\affiliation{Institute of Theoretical Physics, Vienna
    University of Technology, Wiedner Hauptstra\ss e 8-10/136, A-1040
    Vienna, Austria}
\email{svozil@tuwien.ac.at} \homepage[]{http://tph.tuwien.ac.at/~svozil}

\date{\today}

\begin{abstract}
If the unitary quantum mechanical state evolution is universally valid, quantized systems evolve uniformly, deterministically, and reversible; that is, one-to-one. Hence, what is considered an irreversible measurement might be a purely subjective, conventional, and convenient abstraction of the situation that, although in principal totally reversible,  for all practical purposes (fapp), measurements cannot be undone. If this is granted, then Schr\"odinger's ``quantum jellification'' arises because of the inevitability of the physical co-existence of classically mutually exclusive states through quantum coherence. It is suggested to take the rather unique human cognitive and perceptive experience as evidence that, at least at the level of apperception, quantum jellification does not exist at all. Otherwise the problems of how to characterize the ambivalence of perception and cognition induced by quantum coherence on a fundamental level of cognition, and why this ambivalence appears to be rather weak and can be ignored fapp, remain unsolved.
\end{abstract}

\maketitle

\section{Compatibility between the two types of quantum evolutions}
With regard to the evolution of the formal representation of any physical state there appear to be two principal dual quantum {\it modi operandi:}
(i)
on the one hand
quantum mechanics postulates a deterministic, one-to-one evolution of the state inbetween two irreversible measurements;
(ii) yet
on the other hand irreversible measurements are modelled by indeterministic and irreversible, many-to-one mappings.
This duality has already been reflected by Born~\cite[p.~804]{born-26-2}
(English translation in Ref.~\cite[p.~302]{jammer:89}) stating that
{``the motion of particles conforms to the laws of probability, but the probability itself is propagated in accordance with the law of causality.
This means that knowledge of a state in all points in a given time determines the distribution of the state at all later times.''}

At the time of Born's considerations, not too many people seemed to have been concerned about the consequences of these postulates for individual quanta,
let alone their intrinsic consistency.
Alas, with increasing intensity Schr\"odinger worried about the conceptual and empirical ramifications associated with the assumption of a dual quantum evolution.
Early on, in a series of centennial papers~\cite{schrodinger,CambridgeJournals:1737068,CambridgeJournals:2027212}
he pointed out that, if the quantum coexistence of classically distinct and even mutually exclusive states
is taken for granted,
then  due to the very different futures or consequences evolving from such states
-- futures and consequences which can have very different renditions on a macroscopic scale --
seemingly mindboggling if not outrightly absurd consequences follow.
The most famous such example is a cat being in a ``coherent superposition between life and death.''
One may intensify Schr\"odinger's concerns even more by considering this situation not from an external, outside point of view,
but from a quasi-conscious intrinsic~\cite{svozil-94} experience of an individual
suspended in a coherent superposition between life and death.
Alas, as Hilbert once pointedly stated~\cite[p.~163]{hilbert-26},
``the conception that facts and events could contradict themselves appears to me as an exemplar of thoughtlessness.''

Later on, Schr\"odinger addressed related issues by pointing out that
irreversible measurements (quasi conspicuously) appear to ``save the day''
by turning a quantum superposition he termed ``jelly fish'' or ``quagmire'' into a unique phenomenology
\cite[pp.~19--20]{schroedinger-interpretation}:
``The idea that  [[the alternate measurement outcomes]] be not alternatives but {\em all} really happening simultaneously
seems lunatic to [[the quantum theorist]], just {\em impossible.}
He thinks that if the laws of nature took {\em this} form for,
let me say,
a quarter of an hour, we should find our surroundings rapidly turning into a quagmire, a sort of a featureless jelly or plasma,
all contours becoming blurred, we ourselves probably becoming jelly fish.
It is strange that he should believe this.
For I understand he grants that unobserved nature does behave this way -- namely according to the wave equation.
$\ldots$ according to the quantum theorist, nature is prevented from rapid
jellification only by our perceiving or observing it.''

Very similar concerns have been raised by Everett~\cite{everett}, who refers to Process 1 and Process 2
as  ``the discontinuous change brought about by the observation of a quantity,''
and ``the continuous, deterministic change of state of an isolated system with time according to a wave equation,''
respectively.
Everett then continues by considering ``an isolated system consisting of an observer or measuring apparatus, plus an object system.
Can the change with time of the state of the total system be described by Process 2?
If so, then it would appear that no discontinuous probabilistic process like Process 1 can take place.
If not, we are forced to admit that systems which contain observers are not subject to
the same kind of quantum-mechanical description as we admit for all other physical systems.''
And so, pointedly stated, either there is no such thing as an irreversible measurement,
or quantum mechanics is invalid; at least during observations.

Presently the theoretical and empirical findings appear to corroborate the former assumption of the impossibility of an irreversible measurement.
At least in principle, it is commonly believed that any measurement can be undone~\citep{PhysRevD.22.879,PhysRevA.25.2208,greenberger2,Nature351,Zajonc-91,PhysRevA.45.7729,PhysRevLett.73.1223,PhysRevLett.75.3783,hkwz},
although for all practical purposes (fapp) most measurements cannot be undone.
Arguments for a rapid tendency of coherent superpositions to become effectively classical
due to environmental effect resulting in the outflow of correlations and entanglement into uncontrollable regions of spacetime~\cite{RevModPhys.75.715}
contribute to a fapp understanding, but they are of not much help when it comes to questions of principle.

Strictly speaking,  irreversible measurement do not exist.
What is considered a ``measurement,'' and thus the subject-object partition
manifesting itself through the  cut between the measurement tool and the system,
between observer and object, fapp is a convenient approximation, if not an outright illusion.
Any such distinction, though fapp acknowledged, must inevitably be purely conventional~\cite{svozil-2001-convention}.
Hence, from a principle point of view,
it remains totally unclear why we are not living in a jellified universe characterized by Schr\"odinger's quagmire state
consisting of a coherent superposition of all possibilities
and potentialities resulting from the (supposedly random) choices which have simultaneously coexisted and will remain to coexist forever.
Everett's assumption of the simultaneous existence of (conscious) observers which are capable of surfing through this quagmire
by always keeping a classical view of the picture,
thereby multiplying and branching off at every observation (with supposedly random outcome), appears to be a highly questionable
scenario.
For we have not the faintest idea how such a quasiclassical picture could be maintained within the quantum quagmire,
and neither Everett nor anybody else has ever suggested how this might come about.

Another issue is related to multipartite entanglement across spatial distances.
When measuring a concrete particle it cannot be excluded that this particle is not, say,
a partner of a pair of particles in a singlet state; with the other partner ``far away.''
Moreover, interactions between two systems may create a bigger entangled supersystem.
In classical physics measurement of a certain single particle property
in a quasi-singlet  state~\cite{peres222} would be totally justifiable.
This is due to the fact that the states of classical compound systems
can be composed from the states of the individual parts.
This is due to the omniscience a classical observer can at least in principal be  endowed with.
From the principle of quantum complementarity
as well as from Bell-, Kochen-Specker- and Greenberger-Horne-Zeilinger arguments this can no longer be assumed quantum mechanically.
In contrast, the information encoded in a quantum system, through entanglement, may be distributed over some multipartite state;
whereas measurement of the individual constituents would render arbitrary outcomes and might reveal nothing at all about the encoding.
For instance, the singlet state of two spin-half particles is totally characterized by
the joint property that, when measured along two (or more orthogonal) spatial directions,
the outcomes of spin state measurements of the individual particles are exactly opposite~\cite[p.~640]{zeil-99}.
Any outcome of measurements on individual particles reveals no information
encoded originally whatsoever; it is only after (re)combining the individual outcomes that the information
encoded in the joint correlations can be extracted.
It is this property of ``distributive information'' that is at the heart of certain speedups in
quantum computation, such as Deutsch's algorithm.
Thus, in a strict sense, any entangled quantized system shall never be considered to consist of  individual separated parts.
Insisting on and forcing this partial view renders an improper conception of the entangled mind.

\section{Epistemic incompleteness versus phenomenological uniqueness}

In view of the conceptual difficulties discussed so far it might not be totally unreasonable to pursue another, epistemic, route.
Suppose that the wave function is the optimal representation of our intrinsic knowledge
of a quantized system rather than an objective state of the universe; in Schr\"odinger's own words, it is a catalogue of expectations~\cite{schrodinger}.
The representation of a physical system may not be unique, and not even consistent,
because, as expressed by Bohr~\cite{bohr-1949}, it inevitably copes with ``the impossibility of any sharp separation between the behavior of atomic
objects and the interaction with the measuring instruments which serve to define the conditions
under which the phenomena appear.
$\ldots$
Consequently, evidence obtained under different experimental conditions cannot be
comprehended within a single picture, but must be regarded as complementary in the sense
that only the totality of the phenomena exhausts the possible information about the objects.''
Bohr also might have suggested that~\cite[p.~12]{Petersen-63}
``there is no quantum world. There is only an abstract physical description. It is wrong to think that the task of physics is to find out how
nature {\em is.} Physics concerns what we can {\em say} about nature.''

From this point of view, quantum theory might be considered a formalization of Plato's  Allegory of the Cave.
In this scenario,
Schr\"odinger's quagmire could be avoided by acknowledging that, while any physical system  at any particular step of evolution (characterized by time)
is in a single unique state,  fapp this state remains incomprehensible to intrinsic observers.
The optimal representation of this state is given by quantum mechanics,
but as we are inevitably ignorant of the ``true'' state,
any formally ``complete'' representation referring to all potential measurement outcomes  needs to be ambivalent from a classical point of view~\cite{hilbert-26}.
At the same time,
the (at least fapp)  apparent uniqueness of our experience both of the outside world as well as of our self image
indicates that we are not quagmires or jelly fish (hence the title relating to Descartes' Principles of Philosophy),
thereby corroborating the assumptions of a unique but unknowable state of the universe.

\begin{figure}
\begin{center}
\unitlength1.2mm
\thicklines 
\begin{picture}(61.33,36.00)
\multiput(0.33,20.00)(0.36,-0.12){84}{{\color{red}\line(1,0){0.36}}}
\multiput(30.33,10.00)(0.36,0.12){84}{{\color{blue}\line(1,0){0.36}}}
\put(0.33,20.00){{\color{red}\circle{2.50}}}
\put(15.33,15.00){{\color{red}\circle{2.50}}}
\put(30.33,10.00){{\color{red}\circle{2.50}} }
\put(30.33,10.00){{\color{blue}\circle{4.00}} }
\put(45.33,15.00){{\color{blue}\circle{2.50}} }
\put(60.33,20.00){{\color{blue}\circle{2.50}} }
\put(30.33,2.00){\makebox(0,0)[cc]{$(0, 1, 0)$}}
\put(30.33,17.00){\makebox(0,0)[cc]{{\color{red}$\alpha$},{\color{blue}$\delta$}}}
\put(0.33,14.00){\makebox(0,0)[cc]{{\color{red}$\left(-\frac{1}{\sqrt{2}}, 0, \frac{1}{\sqrt{2}}\right)$}}}
\put(0.33,26.00){\makebox(0,0)[cc]{{\color{red}$\gamma$}}}
\put(15.33,8.00){\makebox(0,0)[cc]{{\color{red}$\left(\frac{1}{\sqrt{2}}, 0, \frac{1}{\sqrt{2}}\right)$}}}
\put(15.33,20.00){\makebox(0,0)[cc]{{\color{red}$\beta$}}}
\put(45.33,8.00){\makebox(0,0)[cc]{{\color{blue}$\left(-\frac{i}{\sqrt{2}}, 0, \frac{1}{\sqrt{2}}\right)$}}}
\put(45.33,20.00){\makebox(0,0)[cc]{{\color{blue}$\epsilon$}}}
\put(60.33,14.00){\makebox(0,0)[cc]{{\color{blue}$\left(\frac{i}{\sqrt{2}}, 0, \frac{1}{\sqrt{2}}\right)$}}}
\put(60.33,26.00){\makebox(0,0)[cc]{{\color{blue}$\zeta$}}}
\end{picture}
\end{center}
\caption{(Color online) Diagrammatical representation of two interlinked Kochen-Specker contexts.
\label{2009-context-f1}}
\end{figure}
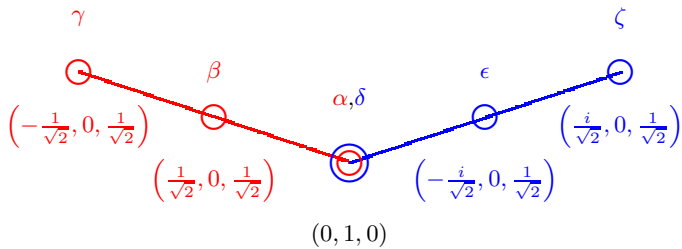

In what follows we shall briefly mention a few phenomenological indications  related to the epistemic approach just mentioned.
Alas as these are neither necessary nor sufficient for the above epistemological assumptions,
they cannot be considered either tests or proofs.
One argument~\cite{svozil:040102} is against quantum contextuality
as exhibited by quantum mechanics itself.
Bell suggested~\cite[p.~451]{bell-66} that
``the result of an observation may reasonably depend $\ldots$ on the complete disposition of the apparatus.''
That is,
the outcome of a measurement of an observable may depend on what other commeasurable observables are measured alongside of that observable.
The simplest nontrivial empirical setup to test this assumption~\cite{hey-red,redhead,svozil-2004-vax,svozil:040102}
is the quantum logic $L_{12}$~\cite{svozil-ql,svozil-2008-ql}
whose structure of propositions is depicted by the Greechie orthogonality diagram~\cite{greechie:71}
(representing orthogonal one-dimensional projectors by smooth, unbroken lines) in Fig.~\ref{2009-context-f1}.

In order to be able to use the type of counterfactual inference employed by an Einstein-Podolsky-Rosen~\cite{epr} setup,
a multipartite quantum state has to be chosen which satisfies the {\em uniqueness property}~\cite{svozil-2006-uniquenessprinciple}
with respect to the two interlinked contexts (or, by another term, subalgebra)
such that knowledge
of a measurement outcome of one particle entails the certainty that, if this observable were measured on the
other  particle(s), the outcome of the measurement would be a unique function of the
outcome of the measurement actually performed.
Consider the two spin-one particle singlet state
$$\left|  \left. \varphi_s \right\rangle  \right. = {1\over \sqrt{3}}\left(-|00\rangle+|-+\rangle+|+-\rangle\right)$$
and identify with the spin states the directions in Hilbert space
$|+\rangle =(1,0,0)$,
$|0\rangle =(0,1,0)$, and
$|-\rangle =(0,0,1)$.
In the Kronecker product representation,
$\left|  \left. \varphi_s \right\rangle  \right. =(1/\sqrt{3})\left(0,0,1,0,-1,0,1,0,0\right)$.
This singlet state is form invariant under spatial rotations (but not under all unitary transformations~\cite{rose})
and  thus satisfies the uniqueness property for all such operations,
just as the ordinary Bell singlet state of two spin one-half quanta.
Hence, it is possible to employ a similar counterfactual argument and establish two elements of physical reality
according to the Einstein-Podolsky-Rosen criterion for the two interlinked contexts (i.e., maximal systems of commeasurable observables~\cite{svozil-ql,svozil-2008-ql})
depicted in Fig.~\ref{2009-context-f1}.

For the sake of modeling observables corresponding to the configuration depicted  in Fig.~\ref{2009-context-f1},
consider
the maximal Kochen and Specker operators~\cite{kochen1} defined by

\begin{widetext}
\begin{equation}
\begin{array}{ccl}
C(\alpha , \beta ,\gamma) &=&  \frac{1}{2}\left[ (\alpha  + \beta - \gamma )J^2(\frac{\pi}{2},0)
+ (\alpha  -  \beta + \gamma) J^2(\frac{\pi}{2},\frac{\pi}{2}) + (\beta + \gamma -\alpha ) J^2(0,0)\right],  \\
C' (\alpha , \beta ,\gamma)&=&  \frac{1}{2}\left[ (\alpha  + \beta - \gamma )J^2(\frac{\pi}{2},\frac{\pi}{4})
+ (\alpha  -  \beta + \gamma) J^2(\frac{\pi}{2},\frac{3\pi}{4}) + (\beta + \gamma -\alpha ) J^2(0,0)\right],\end{array}
\end{equation}
\end{widetext}
for some real $\alpha \neq \beta \neq \gamma \neq \alpha$, where
\begin{equation}
\label{l-soksp}
J(\theta , \phi )=
\left(
\begin{array}{cccc}
\cos \theta & {e^{-i\phi}\sin \theta \over \sqrt{2}}& 0      \\
{e^{i\phi}\sin \theta \over \sqrt{2}}& 0
& {e^{-i\phi}\sin \theta \over \sqrt{2}}      \\
0& {e^{i\phi}\sin \theta \over \sqrt{2}}& -\cos \theta
\end{array}\right)
\end{equation}
stands for the spin one observables (e.g., Refs.~\cite{schiff-55,rose})
in arbitrary directions measured in spherical coordinates.
$0 \le \theta \le \pi$ represents the polar angle in the $x$-$z$-plane taken
from the $z$-axis,
and $0 \le \varphi < 2 \pi$  is the azimuthal angle in the $x$-$y$-plane taken
from the $x$-axis.

The experimentally testable criterion for contextuality can be stated as follows:
Contextuality predicts that there exist outcomes associated with $\alpha$ on one context $C$ which
are accompanied by the outcomes $\epsilon$ or $\zeta$ for the other context $C'$;
likewise $\delta$ should be accompanied by $\beta$ and $\gamma$.
The quantum mechanical expectation values can be obtained from
\begin{widetext}
\begin{equation}
{\rm Tr}\left\{ \bigl| \varphi_s \right\rangle \left\langle \varphi_s  \bigr|
\;\cdot \;
\left[C(\alpha , \beta , \gamma)\otimes C'( \delta ,\epsilon , \zeta )\right]\right\}
=\frac{1}{6} \left[2 \alpha  \delta + (\beta + \gamma) (\epsilon + \zeta) \right]
.
\end{equation}
\end{widetext}
As a consequence, the outcomes
$\alpha$--$\epsilon $,
$\alpha$--$\zeta  $, as well as
$\beta $--$ \delta $ and
$\gamma $--$ \delta $ indicating contextuality do not occur.
This is in contradiction with the quantum contextuality hypothesis.

Another possibility to experimentally corroborate the assumption that quantum mechanics is an optimal epistemic theory
about an inaccessible single and unique ontological state (of the universe and thus of everything physical)
is to test the context translation principle~\cite{svozil-2003-garda,svozil_2010-pc09}
postulating the possibility to reduce, enhance and tune at will
the capacity of a measurement device to translate between the context prepared and the context measured.
Here, again, a quantum mechanical context~\cite{svozil-2008-ql} (or subalgebra of the Hilbert logic)
is a maximal collection of commeasurable observables within the nondistributive structure of quantum propositions.
It can be formalized by a single maximal self-adjoint operator, such that
every collection of mutually compatible commeasurable operators (such as projectors corresponding to yes--no propositions)
are functions thereof~\cite[\S~84]{halmos-vs} (cf. also Ref.~\cite[Sec.~II.10, p. 90]{v-neumann-49}).

\section{Summary}

In summary we have briefly considered three indications or criteria that might be capable of supporting the notion
of a quantum rendition of an essentially  epistemic theory based on an unknowable ``reality.''
The first evidence comes through self-introspection in the sense of Descartes,
by perceiving a rather unique and singular state of the external world,
as well as of the observer's mind with regard to one's own conscious self-experience.
The second argument is based on the independence of single outcomes from the accompanying contexts in  Einstein-Podolsky-Rosen type setups,
and the third suggests to consider the capability of physical systems to translate between a mismatch of preparation and measurement contexts.
The consequences for the applicability of quantum mechanical devices which might be capable for potentially outperforming
classical universal computation, say by
rendering true uncomputability~\cite{2008-cal-svo}, are immense.

\bibliography{svozil}

\end{document}